# Sleep quality, chronotype and social jet lag of adolescents from a population with a very late chronotype

Andrés H. Calderón[1], Romina A. Capellino[1], Damián Dellavale[2], D. Lorena Franco[2], Pablo M. Gleiser[2], Sergio Lindenbaum[1], Mara López-Wortzman[1], Sabrina C. Riva[2], Fernanda R. Román[2], & Sebastián Risau-Gusman[2].

[1] Instituto de Neurociencias Aplicadas, Mitre 265, San Carlos de Bariloche, Argentina.

[2] CONICET and Medical Physics Department, Centro Atómico Bariloche, Bustillo 9500, San Carlos de Bariloche, Argentina.

## Abstract

Sleep disorders can be a negative factor both for learning as for the mental and physical development of adolescents. It has been shown that, in many populations, adolescents tend to have a "poor" sleep quality, and a very "late" chronotype. Furthermore, these features peak at adolescence, in the sense that adults tend to sleep better and have an earlier chronotype. But what happens when we consider adolescents in a population where already adults have poor sleep quality and a very late chronotype? We have conducted two non-clinical studies in the city of Bariloche, Argentina aimed at measuring sleep quality, chronotype, and social jet lag, using the Pittsburgh and Munich questionnaires. These were administered individually to groups of high school students, as well as to smaller samples of adults and preadolescents, in order to study differences between adolescents and these groups. The results show that in this population sleep quality is much poorer than in most other healthy populations recorded elsewhere. Furthermore, sleep quality is consistently worse for adolescents than for the other groups. The difference with adults seems to be due mainly to increased daytime sleepiness and sleep latency, whereas the difference with preadolescents seems to be due mainly to shorter sleep duration.

We also found that the chronotypes of all the groups are very late, with a peak at an age between 18 and 24 ys. Social jet lag and sleep onset latency are also large, and they peak at adolescence, which suggests that they might be closely related to the large prevalence of poor sleep quality that we find in adolescents.

## Introduction

Sleep disorders can be a negative factor both for learning as for mental and physical development, specially in the adolescent population (1, 2). In particular, they can lead to daytime sleepiness which in turn affects driving, working conditions and school performance, and may produce irritability, anxiety, and depressive moods.

In adolescents, chronic insufficient sleep has been shown to cause a decrease in their ability to understand the consequences of risky behaviors (3), and to increase susceptibility to peer pressure and even to develop suicidal thoughts (4). Poor sleep is thus an important public health concern, and an assessment of its causes and features at a population level is a necessary first step towards a resolution of this issue.

Actigraphy and polysomnography are the most direct methods to study sleep disorders, but their complexity makes them rather impractical for population studies. Besides, it is not clear that they are the best methods to assess sleep "quality" (5), a concept that has a strong subjective component in its very definition. Even though necessarily less direct (6), questionnaire-based method provide a more practical approach to assessing the sleep quality of large groups.

One of the most widely used instruments is the Pittsburgh Sleep Quality Index (PSQI), a questionnaire-based test designed by D. J. Buysse (7) that has since been validated and used in many populations and in several languages. It was designed by taking into account psychiatric clinical experience as well as theoretical knowledge about sleep quality and disorders. PSQI analyses sleep by measuring seven components: sleep quality, sleep latency, sleep duration, habitual sleep efficiency, sleep disturbance, use of sleeping medication and daytime dysfunction. A number of studies have confirmed the high internal homogeneity, internal consistency and reliability of each component (8). The PSQI has been widely used in several epidemiological studies across the globe, including countries such as Germany (9), Austria (10), Brazil (11), and Hong-Kong (12). Depending on the region, poor sleep quality rates can differ considerably in the adult general population: 36% in Germany (9), 39.4% in Hong-Kong (12), 32.1% in Austria (10), and more than 50% in Chile, Ethiopia and Peru(8). In adults, women tend to have a poorer sleep quality than men, independently of the geographic area (9, 10, 12, 13). Furthermore, poor sleep quality appears to increase with age (9, 10, 12).

It is by now well known that sleep quality can be affected by natural, as well as by social factors. One of the functions of the endogenous circadian clock, present in most animals, is to control activity-rest periods (14). However, there are large differences among individuals in the way that sleep-wake patterns are controlled (15), which gives rise to the concept of chronotypes (16). People with a late chronotype ("owls") tend to go to bed late and wake up late, whereas people with an early chronotype ("larks") tend to go to bed early and wake up early. During adolescence, chronotype becomes increasingly late, and some studies show that it seems to reach a maximum at an age between 18 and 20 (17, 18). It has even been suggested that this maximum could be considered as a good marker for the end of adolescence (19). Another very related variable that peaks in adolescence is "social jet lag" (20). This is a measure of the mismatch between the "social" and the "biological" clock, and is quantified as the difference between the average sleep hours on workdays (weekdays in the case of adolescents) and on weekends. In adolescents jet lag is particularly large because naturally late bedtimes are combined, on weekdays, with early start times of school.

Even though adolescent sleep has been extensively studied (4, 21, 22, 23, 24, 25), measurements of sleep quality using the PSQI in adolescents have been much less frequent. In general it has been shown that adolescents have both poor sleep quality and a relatively late chronotype, but in many cases the comparison with adults is made between results from different studies. In order to have a clear picture of the age dependence of sleep quality, it is important that the study is conducted at the same time in all the populations that are to be compared. Given that sleep habits are very dependent on culture, it is important to study adolescent sleep in population from different geographies and cultural contexts whose sleep habits could differ from what is usually considered "normal". In particular, one can ask whether sleep quality is still poorer for adolescents in populations where sleep quality is already very poor in general. These are the questions that we have tried to address in this article.

In urban areas of Argentina sleep habits seem to be strongly influenced by cultural ties with Spain, with bedtime taking place much later than in other countries. Some studies have shown that, at least for some groups in Argentina, sleep quality tends to be rather poor. Furthermore, it has recently been shown that in some populations there is an effect of latitudinal cline in some of the features that negatively affect sleep (26, 27, 28). For example, in a study carried out in Chile, it has been shown that differences between duration of sleep in workdays and weekends becomes larger in more southern latitudes (26), leading to a larger social jet lag. Thus, it is reasonable to anticipate that in urban areas of southern Argentina sleep quality can be particularly poor. The study presented here was conducted in the city of Bariloche, in southwestern Argentina. In the first part we measured sleep quality in a non-clinical sample of adolescents, adults and children. Our aim was to quantify the differences in sleep quality between adolescents and adults on one side, and between adolescents and children on the other. As sleep quality was found to be consistently poor for all ages, but particularly for adolescents, we also wanted to understand what were the determinants of these differences. In the second part of this study we sought to determine the chronotype and social jet lag in the same population in order to investigate the possible correlations between these features and sleep quality.

## Materials and methods

*Study samples*

The first part of this study, aimed at assessing the sleep quality of an adolescent population and comparing it with sleep quality of other age groups, was conducted in the city of Bariloche, Argentina, as part of the activities of the *Semana del Cerebro* (March, 2017), the local version of the international *Brain Awareness Week*. The activities were oriented mainly to high school students, and also included a number of visitors from the general population of Bariloche. We constrained our analysis to individuals under 50ys, because older adults were very underrepresented in our sample. The groups we considered were: preadolescents between 12 and 14 ys, adolescents between 15 and 17 ys, and adults between 18 and 49 ys. Adolescents were further divided into three groups (15,16 ,17). whereas adults were divided into two groups (18-24, 25-49). The adult groups comprise individuals that do not attend school, whereas all the individuals of the other groups attend school. The group of adults between 18 and 24 years old was composed mainly by university students. The characteristics of the six resulting groups are given in Table 1.

The second part of this study took place in the 2018 edition of the same event. The sample was somewhat smaller than in the first study, but its composition, in terms of socio economical status and geographic origin was very similar. The reason for this is that most of the schools that attended the event in 2017 also attended it in 2018. For statistical analysis the sample was divided into 5 groups: adolescents 15, 16 and 17 years old, adults between 18 and 24 years old ("young adults"), and adults between 25 and 49 years old. The only differences with the grouping performed with the sample of the first study is that in this case preadolescents are not considered (their number was too low). The characteristics of the groups considered in the second study are given in Table 2.

[Table 1 near here]

[Table 2 near here]

*Pittsburgh Sleep Quality Index*

In order to assess sleep quality we used the Pittsburgh Sleep Quality Index (PSQI)(7), obtained from a questionnaire administered individually. Questionnaires with unanswered items, more than one answer in multiple choice items, and other inconsistencies, were not considered. The questionnaire we used was a translation into Spanish of the original English language questionnaire. It was specifically adapted to the version of Spanish that is spoken in Argentina, and it consists of 18 questions, whose scores are used to compute the values of 7 components. These components are: C1-Subjective Sleep Quality (sleep quality as perceived by the subject), C2-Sleep Latency (time lapse subjectively perceived between lying in bed and falling asleep), C3-Duration of Sleep (time spent sleeping), C4-Sleep Efficiency (time spent effectively sleeping of the total time spent in bed), C5-Sleep Disturbances (interrupting disturbances occurred during sleep time, such as heat, snoring, etc.), C6-Use of sleep medication (drug use, with or without medical prescription, to improve sleep quality), C7-Daytime dysfunction (dysfunctions caused by poor sleep quality). The possible components scores are 0, 1, 2 and 3, with 0 and 3 denoting the lowest and highest contributions to sleep quality from the feature assessed by the component. The PSQI score, ranging from 0 (best sleep quality) to 21 (worst sleep quality) is obtained by summing up the scores of the 7 components. For our sample the internal consistency of the questionnaire was acceptable (Cronbach alpha=0.6).

## Chronotype and social jet lag

In order to assess chronotype and social jet lag we used the Munich ChronoType Questionnaire (29). The version we used was a translation into Spanish of the original English language questionnaire, and was specifically adapted to the version of Spanish that is spoken in Argentina. The questionnaire consists of two sets of eight questions where respondents report on their typical sleep behavior over the past 4 weeks both during work (or school) days and during free days. Questionnaires with unanswered items were not considered. For each respondent the answers give the time at which the person falls asleep and the time when the person wakes up. The total duration of sleep is computed as the difference between these two times for workdays and free days (SDw and SDf), whereas midsleep is computed as the midpoint between the two times for workdays and free days (MSw and MSf). Social jet lag is then computed as MSf-Msw. Chronotype is computed as MSf, if Sdf <= SDw, and as MSf - (SDf - SDw)/2, if Sdf > SDw.

## Procedure

The event where both studies were conducted (in 2017 and 2018) consisted of four stands with interactive activities about different aspects of neuroscience. One of the stands was dedicated to sleep (sleep quality in 2017 and chronotype in 2018). Several high school groups visited this stand during the week of the event, and the questionnaire was administered in this context to every member of the group (students and teachers) as well as to occasional visitors. Participants were always assisted by a member of the staff (composed by researchers and medical professionals) and had approximately 15 minutes to complete the questionnaire. The statistical analyses were performed after the event.

All statistical analyses were performed using R version 3.0.2 (R Development Core Team, 2013). To study gender-specific and sex-specific differences in sleep variables, two-way ANOVAs analyses were carried out, followed by Mann-Whitney pairwise comparisons. The level of significance used was 0.05, adjusted by a Bonferroni correction.

*Ethical aspects*

This study was approved by the ethics committee of the Balseiro Institute, National University of Cuyo, Argentina. Oral informed consent was required to the academic tutors of the student groups and informed assent was obtained from all the participants.

## Results

*Mean PSQI is larger for women than for men*

In Fig.1 we compare the distributions of PSQI values of females and males. The proportion of good sleepers (PSQI <=5) is 24.9% for females and 29.5% for males and there is no statistical difference between these values (chi squared=1.937, p=0.164). On the other hand, mean PSQI is significantly larger for women (7.92) than for men (7.16) (Wilcoxon rank sum test, z=-3.14, p=0.0008).

[Figure 1 near here]

*Quality of sleep is poor for all groups and poorest for adolescents*

Fig.2 shows, for each age group, two different (but obviously related) ways of assessing sleep quality. The proportion of individuals of each group that have a PSQI>5 ("bad sleepers"), depicted in the left panel, has a clear maximum at ages 16 and 17, and this maximum is significantly larger than the proportion of bad sleepers in the group of adults and in the group of preadolescents. The mean value of the PSQI for each group is shown in the right panel of Fig.2. This quantity also has a maximum for groups 16 ys and 17ys.

When the sample is divided by gender, in both cases there is a maximum in adolescence (at 15ys and 16ys in men, and at 16ys and 17ys in women), but the differences between groups are more marked in women (see Fig.3). In fact, the differences in mean PSQI between the 16ys and 17ys groups and the

12-14 group are significant (p=0.01 and p=0.02). The differences between genders become more marked after the age of 15 ys but, because of the size of the groups, our statistical tests do not have the power to attribute significance to these differences. In order to obtain a more detailed view of the characteristics of adolescent sleep, it is necessary to perform a separate analysis of the components that form the PSQI.

[Figure 2 near here]

[Figure 3 near here]

*Main determinants of poor sleep*

As happens in many other cases (e.g. 7, 9, 13), we have found that the contribution of component C6 ("use of sleeping medication") to the PSQI is almost negligible. Only 5.9% of the participants (44 out of 739 individuals) reported using such medication regularly. For this reason this component is not included in the analysis that follows.

A comparison between the panels of Fig.4 shows that, for each age group the contributions of each component to the mean PSQI are not very different, as was to be expected. For all age groups, the largest contribution to the mean PSQI is given by component 2 (between 19% and 21%), whereas the smallest contribution is given by component 4 (between 9% and 12%).

Fig.4 also shows that 4 out of the 6 components considered have a significant age dependence. These components are: C1 ("Subjective sleep quality"), C2 ("Sleep latency"), C3 ("Sleep duration") and C7 ("Daytime dysfunction"). Interestingly, in all 4 cases the maximum is located in the groups of 16 and 17 ys, just as in the case of PSQI (Fig.3). In terms of age, the largest differences appear at components C3 and C7: adolescents sleep significantly less than preadolescents ($p<0.03$), and suffer significantly more from daytime dysfunctions than adults ($p<0.007$).

Statistical analysis of the component scores reveals that only for C4 and C5 there is a significant difference between genders. As happens with the PSQI, the scores of these components are larger for females than for males for all individuals of age>15 ys. In fact, the same can be observed in all the other components, but the difference is less marked.

For each component the difference in score across age groups is much more pronounced for females than for males, with the exception of C7, where significant differences between scores appear for both males and females between adolescents and adults (see Fig.5). The significant differences that we find between female adolescents and preadolescents in C1 and C3, are very likely to be important for males too, but, because the size of our sample is too small, the tests we use do not have enough statistical power to detect these differences.

[Figure 4 near here]

[Figure 5 near here]

*Correlation between PSQI and its components*

The correlations between each component and the PSQI are presented in Table 3. All the components have similar, moderate correlations with PSQI, which is an indication that all of them provide information that is important to assess sleep quality. C3 ("Sleep duration") is the component that has the largest correlation with PSQI, for all age groups. Interestingly, even though C7 ("Daytime dysfunctions") is responsible for a large part of the difference in mean PSQI between adolescents and adults, it is the component that shows one of the smallest correlation for all ages. This means that some people with low PSQI report large daytime dysfunctions, but also that some people with relatively high values of PSQI report low daytime dysfunctions (see Fig.6). In turn, this implies that the correlation between C7 and the other components is also rather low.

Component C1 gives information on the sleep quality as perceived by the subjects. The relatively high correlation between this component and the PSQI shows that the score obtained by the individuals tends to be in good agreement with their own perception of sleep quality (see Fig.7).

[Table 3 near here]

[Figure 6 near here]

[Figure 7 near here]

*Chronotype and social jetlag*

In the second part of our study we studied the gender and age dependence of chronotype and social jet lag in a different sample of the same population. We found no significant differences in chronotype (Welch ANOVA, p=0.17) nor in social jet lag (p=0.24) between females and males. On the other hand, we did find a significant age dependence for both chronotype and social jet lag ($p<10^{-11}$). Fig.8 shows that whereas chronotype has a significant maximum for young adults (18-24 ys.), social jet lag is significantly larger for adolescents (15 and 16 ys.) than for young adults and adults (18-24 and 25-49 ys). Even though social jet lag, that measures the differences in sleep duration between work days and free days, is an important indicator of sleep problems, sleep durations by themselves can also give useful information (see Fig.9). In our case, it shows that adolescents sleep very little on week days.

In order to understand the relationship between chronotype and social jet lag it is useful to consider explicitly the times of sleep onset (i.e. the time at which sleep effectively begins in each individual) and wake up, since both chronotype and social jet lag are obtained from them. This information is shown in Fig.10, for each group, discriminating between work days and free days (school days and weekends, respectively, in the case of adolescents). The vertical distance between the corresponding boxplots gives a good idea of the duration of sleep in each case.

Fig.10 shows that the main contribution to social jet lag is the difference in wake up times between weekends and schooldays. The very small variation (i.e. small boxplots) in wake up times in school days is evidently related to the start time of secondary school (between 7.30 and 8.00 in almost all cases). It is interesting to see that when children grow, and this "restriction" is lifted, it is only this wake up time that undergoes a significant change (compare the 18-24 group with the 15, 16 and 17 ys groups in Fig. 10). Note also that the chronotype maximum at 18-24 ys is caused by a maximum in all four times involved.

Another important variable that can be calculated from the answers of the Munich questionnaire is sleep onset latency (SOL), defined as the amount of time between bedtime and the onset of sleep. The values obtained in our second study are very large, as Table 4 shows.

[Figure 8 near here]

[Figure 9 near here]

[Figure 10 near here]

[Table 4 near here]

## Discussion

One of the results of our first study is the fact that sleep quality is poor for all ages in the population studied, and the mean PSQI is among the highest values reported in the literature for otherwise healthy groups. For adults, similar values were reported for Argentina in some studies for some specific populations such as slum dwellers (30) and short distance drivers (31), but our study suggests that poor sleep could also be a problem for the general population in Argentina. The difference with other countries could be a reflection of geographical variation of sleep habits (32).

The fact that in adolescence quality of sleep is "poor" is very well known and has been confirmed in almost every study of adolescent sleep. In the cases where the PSQI is used, "poor" is quantified as a PSQI value larger than 5. However, for this to be meaningful it is also necessary to know how prevalent poor sleep is in the population where the adolescents live. This information is sometimes taken from other studies, but this can be somewhat misleading, since it is known that variability between different studies is very large (33). Direct comparisons are more meaningful, but in general adolescents are either compared with adults or with preadolescents. Here we have shown a direct comparison between adolescents and both adults and preadolescents. We found that, even in a population whose sleep quality is in general very poor, there is a peak at adolescence, and that the differences, even if not very large, are significant. This peak is more marked for women than for men, which is consistent with the fact that pubertal development is associated with an increase of sleep disorders in women, but not in men (34).

Poor sleep quality in the group of young adults (18-24 ys. old) is probably due to the fact that most individuals in this group were university students, who usually have sleep problems, as has been shown in many studies in many countries (see e.g. 35, 36, 37, 38).

The main differences in sleep quality between adolescents and adults are in sleep duration and subjective sleep quality. We found that differences in duration are almost exclusively due to later bedtimes of adolescents (no difference in latencies nor rise times), which is the same pattern that was found in studies of school preadolescents in the US on weekdays (1). This is further indication that, because questions of the Pittsburgh Questionnaire do not discriminate between sleep habits on weekdays or weekends, people tend to choose to report their behaviour on weekdays (39).

Daytime sleepiness has long been recognized as one of the more frequent (and serious) sleep disorders in adolescence (40, 41, 42). In our study we have found that, of the seven factors that compose the PSQI, the one that accounts for most of the difference in sleep quality between adults and adolescents is daytime dysfunction, which usually refers to daytime sleepiness. The correlation between this component and the PSQI was very low, which in turn implies that the correlation with the other six components was almost negligible. It is usually assumed that the main cause of sleepiness in adolescents is lack of sleep, but it has also been suggested that it may be a consequence of brain maturation (43, 44). Our results seem to support the later hypothesis since the correlation between daytime dysfunction and sleep duration was very low (this low correlation is not uncommon in PSQI studies, 5). Nevertheless, it must be noted than in our case virtually all adolescents reported sleeping much less than they needed. In our first study the mean duration of sleep reported was 6.47 hs (sd=1.37), which is very far from the average of 9 hs that is usually recommended (45, 46). But, as mentioned above, it has been argued that the Pittsburgh questionnaire is more correlated with behavior on weekdays. In our second study, however, we found that even during weekends more than half of the adolescents sleep less than the recommended 9 hs.

The chronotypes present in the population that we have studied are particularly late when compared with other populations around the globe, specially in the case of adolescents. Similar results were obtained in a study of chronotype in students in German schools around the world (47). The Argentine students were almost the latest chronotypes (only Spanish students had later chronotypes) with a value of 6.24, very similar to the values reported here (47). The age dependence that we found for chronotypes and social jet lag agree well with results obtained in European populations (15). In particular, the maximum of chronotype found here happens in the group of 18-24 ys, whereas the maximum in social jet lag happens before this age, which is consistent with the findings of other studies (38).

In our study the age at which sleep quality is worst coincides with the age at which social jet lag is largest. This suggests that chronotype only affects sleep quality through social jet lag. Interestingly, the same happens for other sleep-deprived groups (48). It must be noted however, that our study also shows that daytime sleepiness (probably produced by insufficient sleep on weekdays) is almost completely uncorrelated with many other features that are also responsible of poor sleep quality. These other features could in principle also be related to chronotype.

The bedtimes of adolescents in our study are very late, but they are similar to bedtimes found in other parts of the world (49). What is remarkably different is the sleep onset latency (SOL). For the adolescents in our sample, the mean SOL was 81 m (sd=60m), which is very large when compared with other estimates (for example, in a study carried out in Norway a mean value of SOL = 47 m was considered large, 50). A long SOL during prolonged periods of time is one of the defining characteristics of insomnia (51). For adults, the threshold is 31 minutes, but it has been argued that for adolescents this threshold should be 60 minutes, due to the important biological changes the take place during this developmental stage (52). In our study we find that SOL is larger than this threshold for 56.5% of the adolescents (in the Norwegian study this happens for 33.6%). Thus, this large value of SOL is very likely to be related to the very poor sleep quality that we found in our study. This could also be an important factor to explain the large prevalence of poor sleep quality in our sample of adults, since for 61.8% of adults SOL is larger than 31 m.

We have shown here that PSQI is very correlated with the individuals perception of its own sleep quality (which is component 1 of the PSQI). Given that, as mentioned above, PSQI is more correlated with sleep on workdays (or weekdays) (25), it would be interesting to add some questions to the questionnaire in order to have a more thorough understanding of sleep quality and its relation with its subjective perception. This would be particularly important for adolescent studies, since their sleep is very different in weekdays and in weekends. An extension of our work would be to study simultaneously chronotype and PSQI in a population that includes preadolescents, adolescents and adults, in a more controlled setting and with more subjects. Studies of the adolescent poor sleep quality peak in other populations would be useful to understand how widespread are the differences between adolescent sleep on one side, and adult and preadolescents sleep on the other.

Further studies would also be necessary to determine whether the large values of SOL that we find, that are larger than what has been found in other regions, are a particular feature of the Argentinian population. It could also happen that this difference is merely related to the increase in the use of electronic devices before sleep in the last 5 years, since it has been shown (52) that the use of such devices increases the probability of long SOLs.

## Acknowledgments

We want to thank the many volunteers of *Semana del Cerebro* in Bariloche, whose collaboration was essential to administer both the Pittsburgh and the Munich questionnaires. We also want to thank all the students, teachers and occasional visitors who took part in these studies.

## Disclosure Statement

Financial disclosure: None.

Non-financial disclosure: None.

**Table 1: Groups used in the first study**

| Age(ys) | M | F | All | Group |
|---|---|---|---|---|
| 12-14 | 86 (71.1%) | 35 (28.9%) | 121 | preadolescents |
| 15 | 46 (43.4%) | 60 (56.6%) | 106 | adolescents |
| 16 | 69 (44.2%) | 87 (55.8%) | 156 | adolescents |
| 17 | 64 (43.2%) | 84 (56.8%) | 148 | adolescents |
| 18-24 | 34 (31.8%) | 73 (68.2%) | 107 | young adults |
| 25-49 | 31 (30.7) | 70 (69.3%) | 101 | adults |
| All | 329 (44.5%) | 410 (55.5%) | 739 | |

*Table 1. Age and gender of groups participating in the first part of the study (sleep quality).*

**Table 2: Groups used in the second study**

| Age (ys) | M | F | All | Group |
|---|---|---|---|---|
| 15 | 48 (52.7%) | 43 (47.3%) | 91 | adolescents |
| 16 | 36 (35%) | 67 (65%) | 103 | adolescents |
| 17 | 42 (43.8%) | 54 (56.2%) | 96 | adolescents |
| 18-24 | 35 (39.3%) | 54 (60.7%) | 88 | young adults |
| 25-49 | 16 (29.1%) | 39 (70.9%) | 55 | adults |
| All | 177 (40.8%) | 257 (59.2%) | 434 | |

Table 2. Age and gender of groups participating in the second part of the study (chronotype and social jet lag).

**Table 3. Correlations between PSQI and its components.**

| Comp. | All ages | 12-14 | 15 | 16 | 17 | 18-24 | 25-49 |
|---|---|---|---|---|---|---|---|
| C1 | 0.61 | 0.54 | 0.61 | 0.60 | 0.59 | 0.61 | 0.71 |
| C2 | 0.61 | 0.57 | 0.61 | 0.59 | 0.62 | 0.61 | 0.64 |
| C3 | 0.67 | 0.70 | 0.73 | 0.65 | 0.69 | 0.65 | 0.63 |
| C4 | 0.59 | 0.62 | 0.55 | 0.52 | 0.53 | 0.69 | 0.65 |
| C5 | 0.47 | 0.38 | 0.53 | 0.44 | 0.46 | 0.53 | 0.57 |
| C7 | 0.49 | 0.48 | 0.56 | 0.43 | 0.48 | 0.50 | 0.49 |

*Table 3. Correlations between PSQI and each of the components that compose it, for all age groups of the first study.*

**Table 4. Sleep onset latency for all age groups.**

| SOL | | 15 | 16 | 17 | 18-24 | 25-49 |
|---|---|---|---|---|---|---|
| wd | mean(sd) | 82(57) | 72(56) | 89(67) | 88(67) | 50(37) |
| | median | 70 | 65 | 70 | 67 | 45 |
| fd | mean(sd) | 103(77) | 83(80) | 85(71) | 87(68) | 44(34) |
| | median | 90 | 65 | 65 | 70 | 40 |

*Table 4. Sleep onset latency (SOL) in minutes for all age groups of the second study. wd: work (school) days. fd: free days (weekends).*

**Figure 1.**

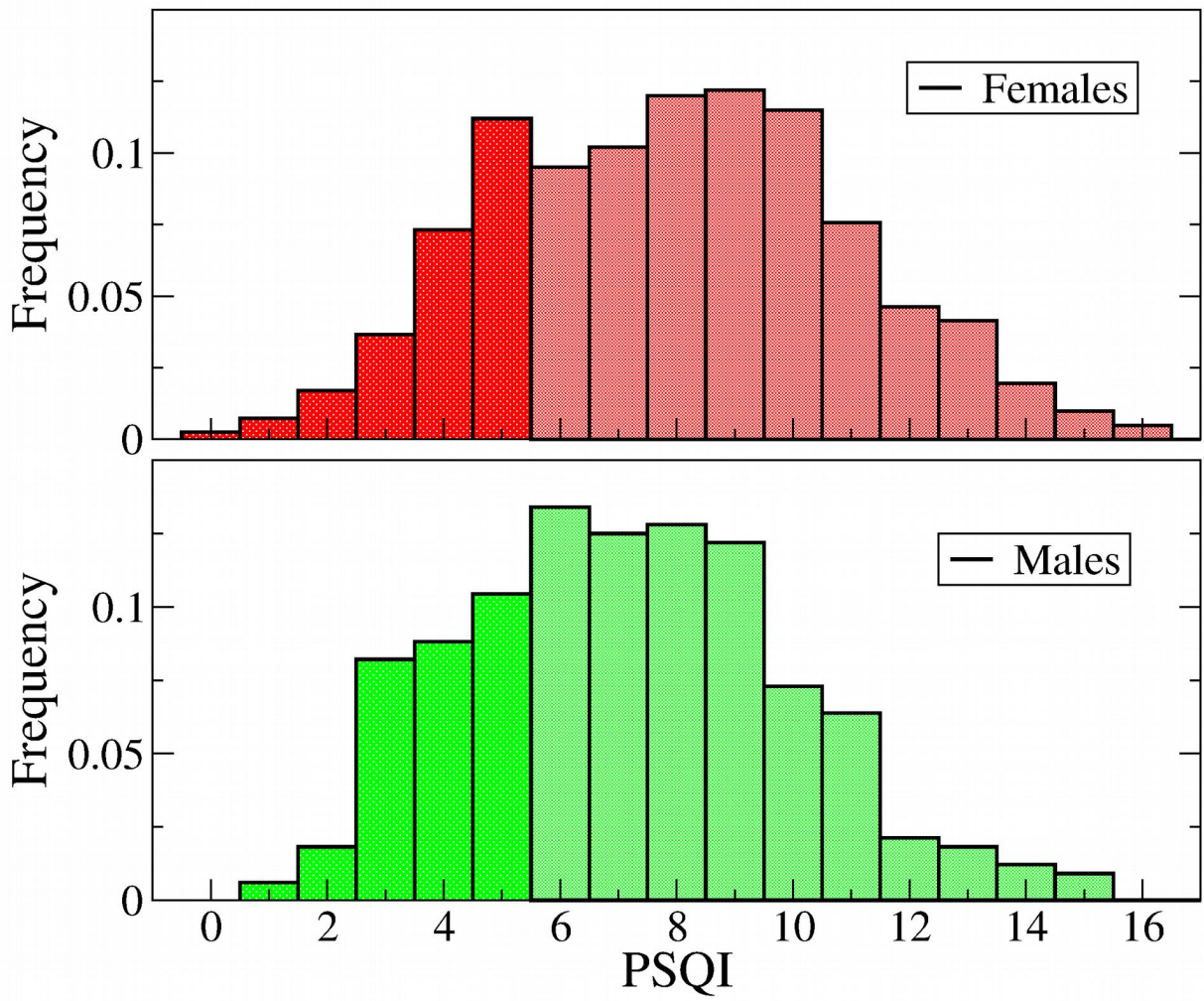

**Figure 2.**

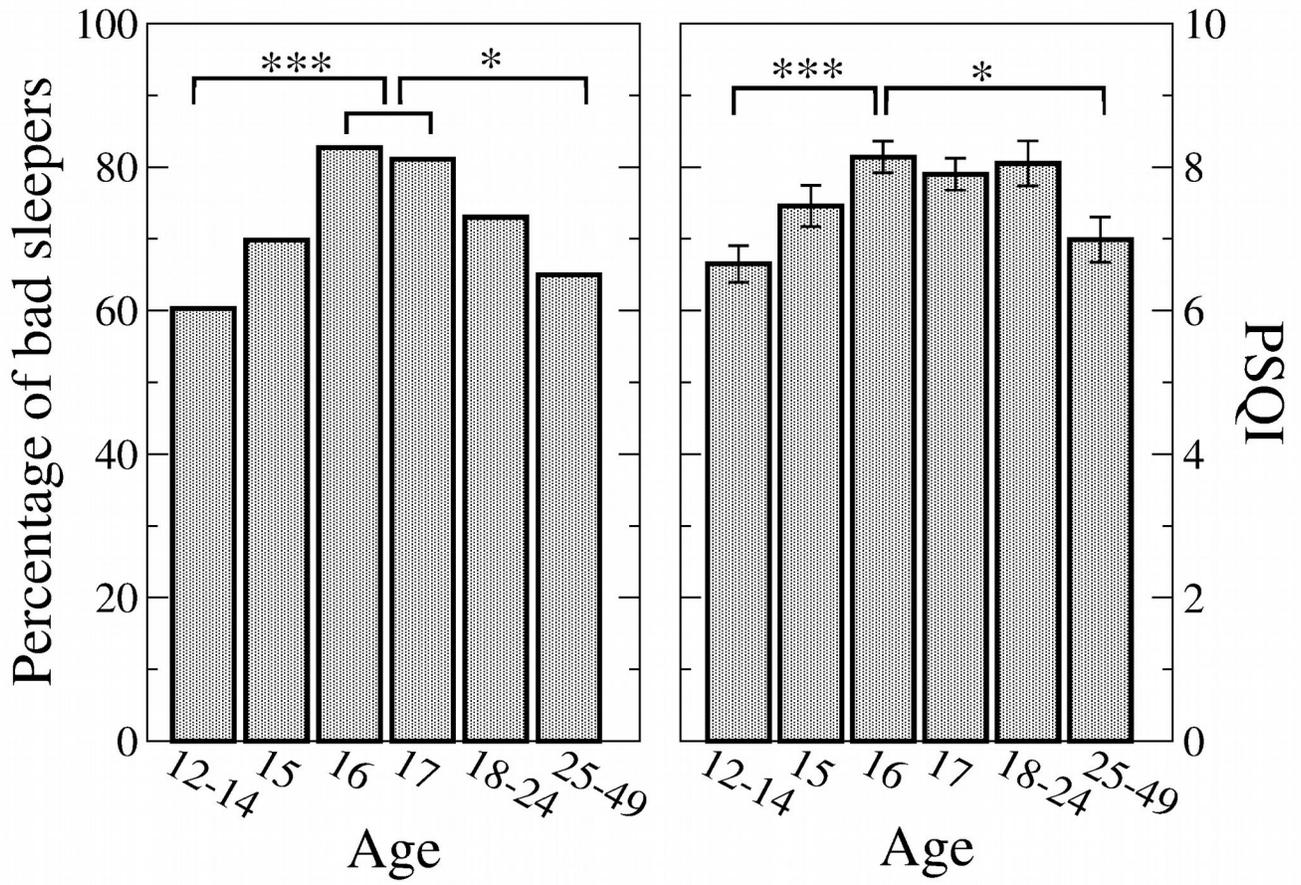

**Figure 3.**

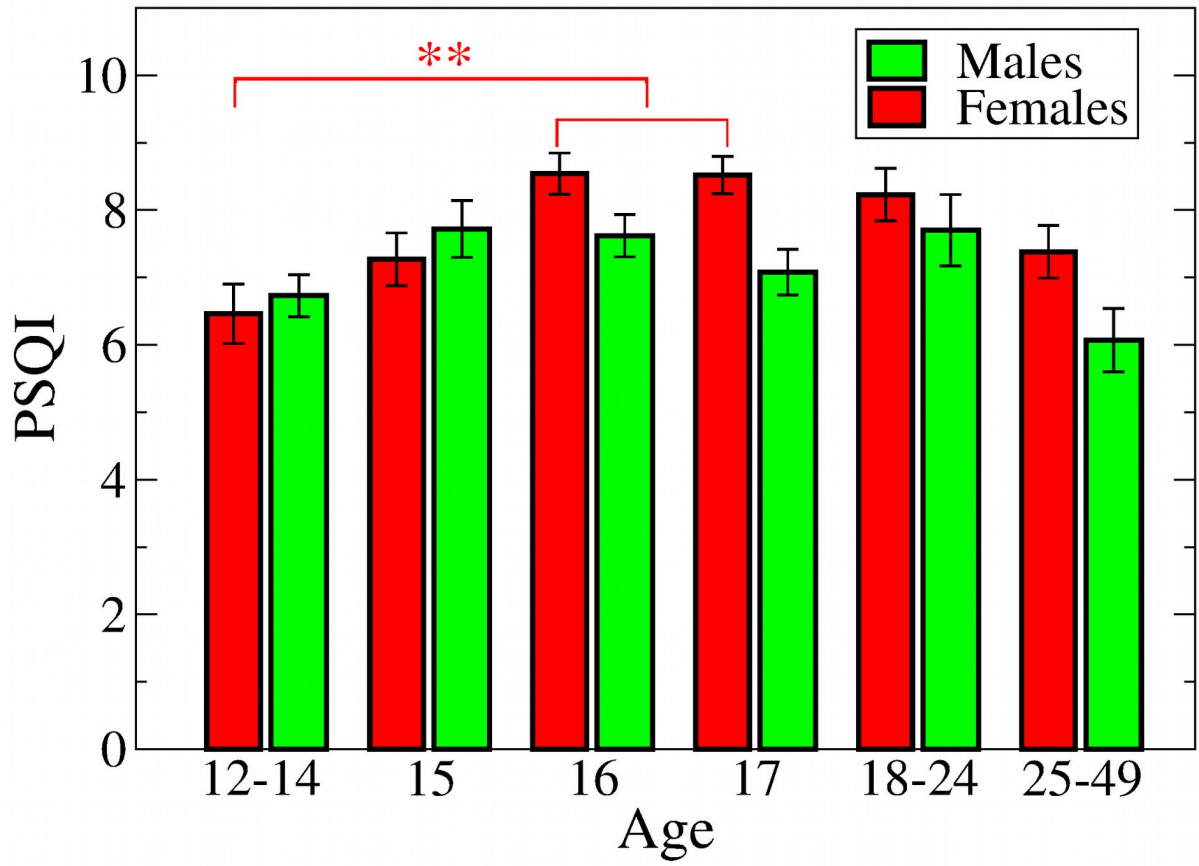

**Figure 4.**

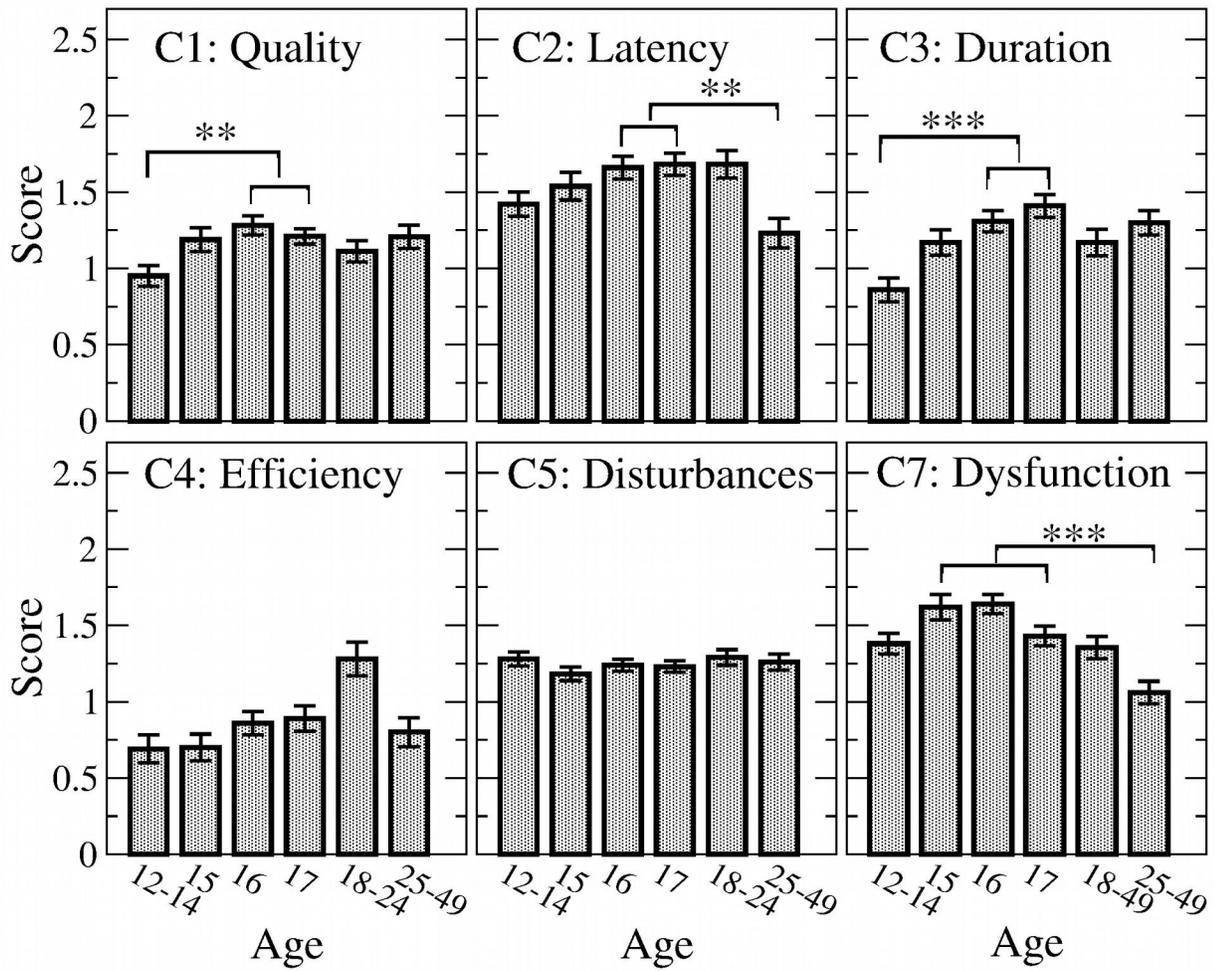

**Figure 5.**

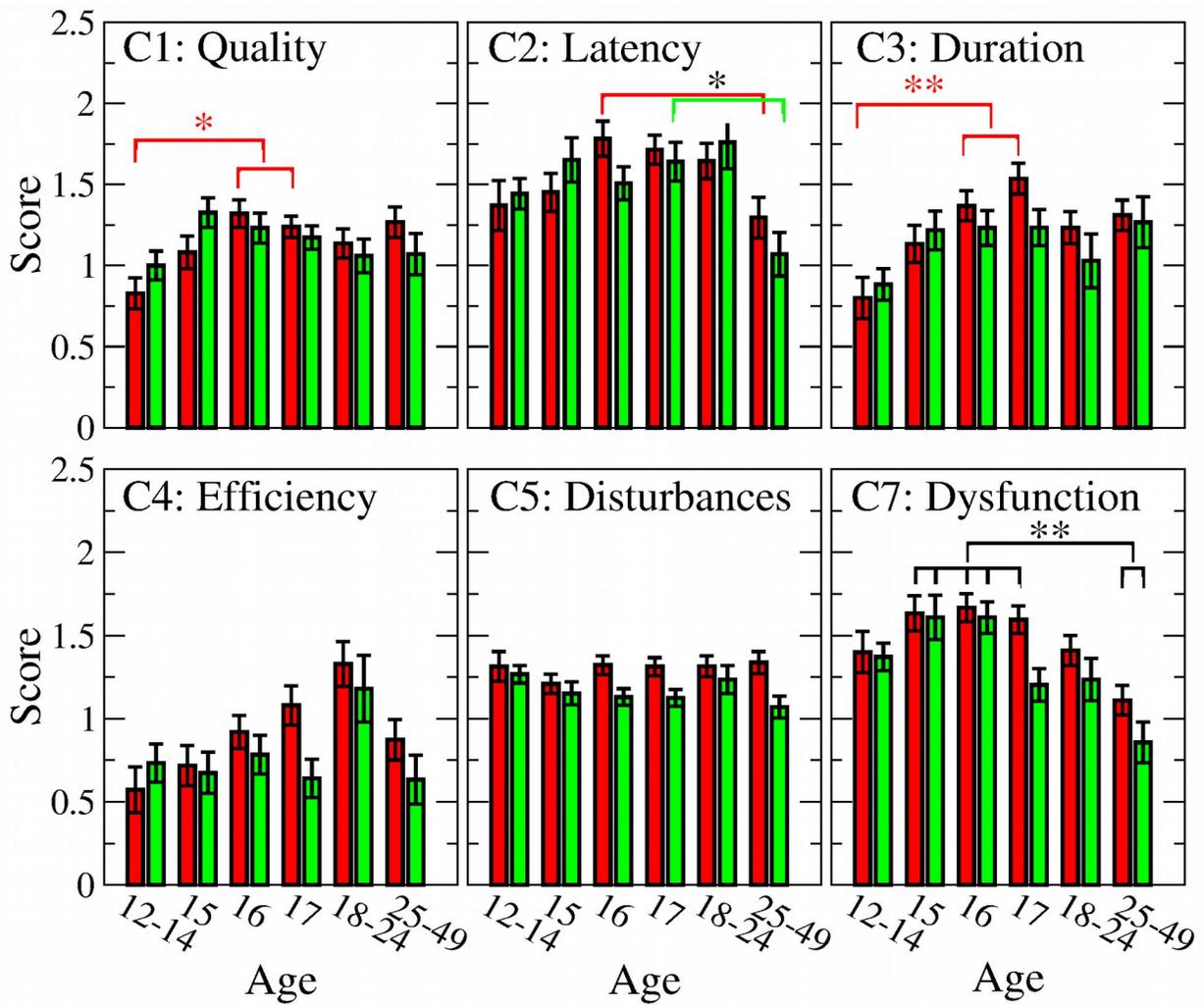

**Figure 6.**

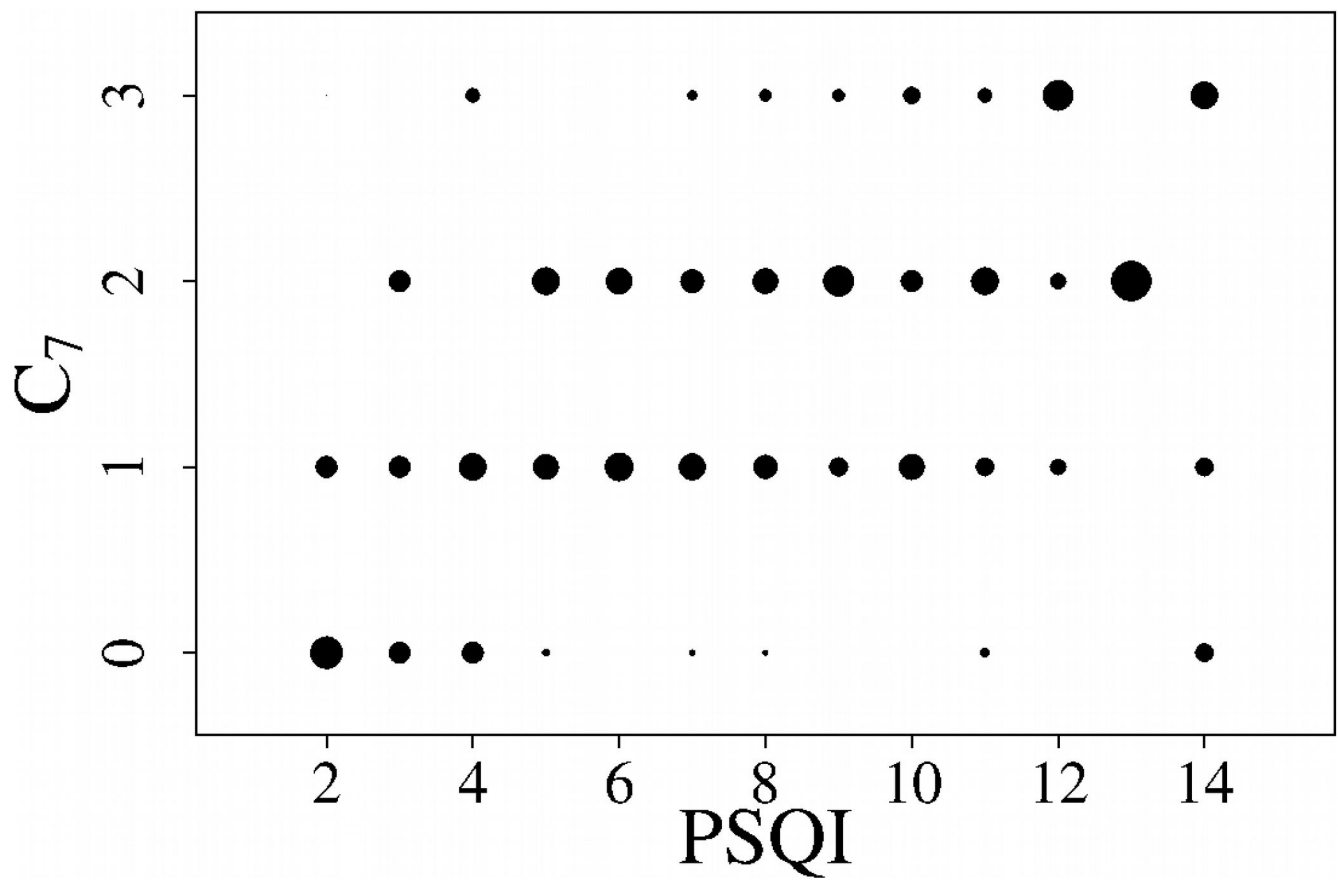

**Figure 7.**

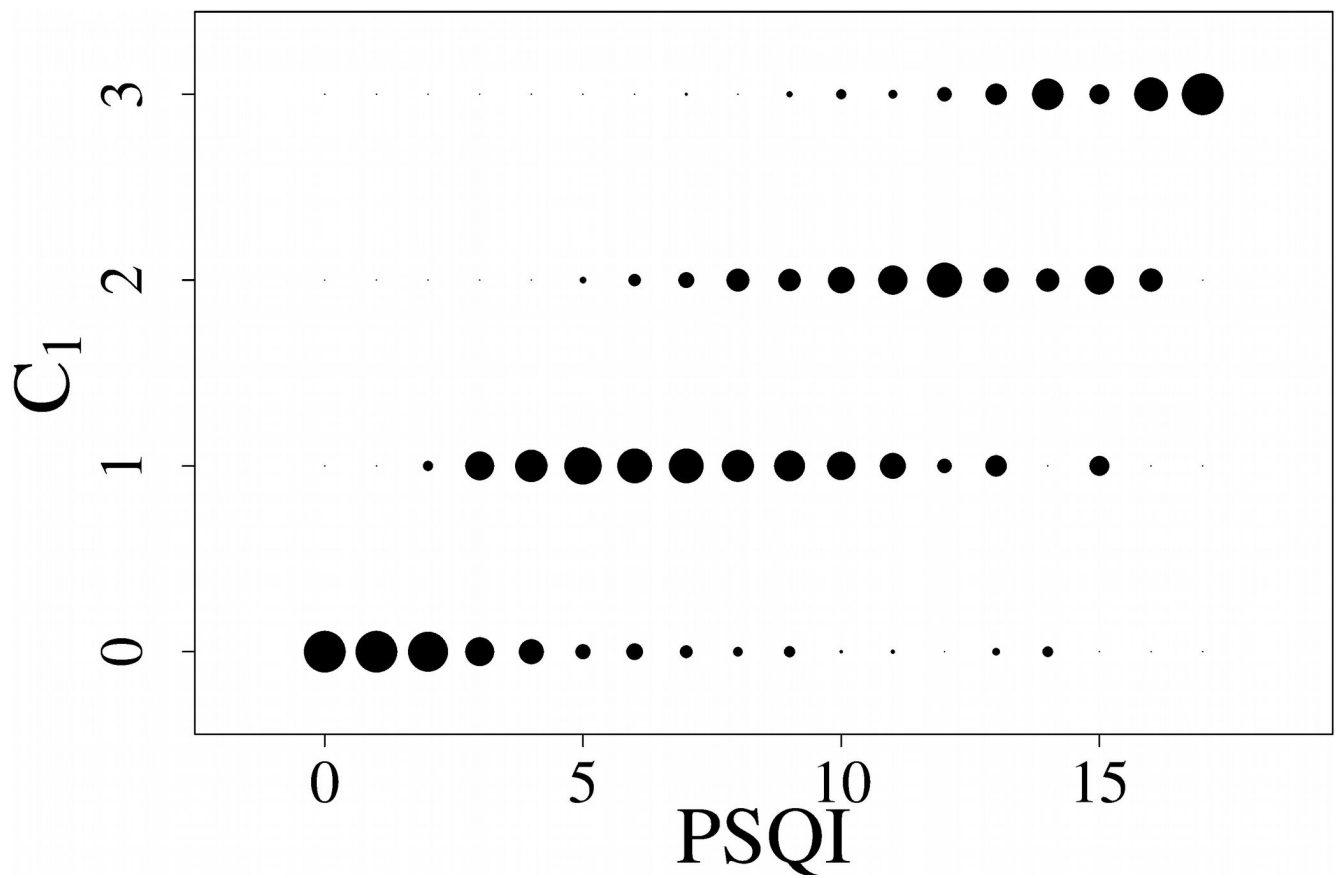

**Figure 8.**

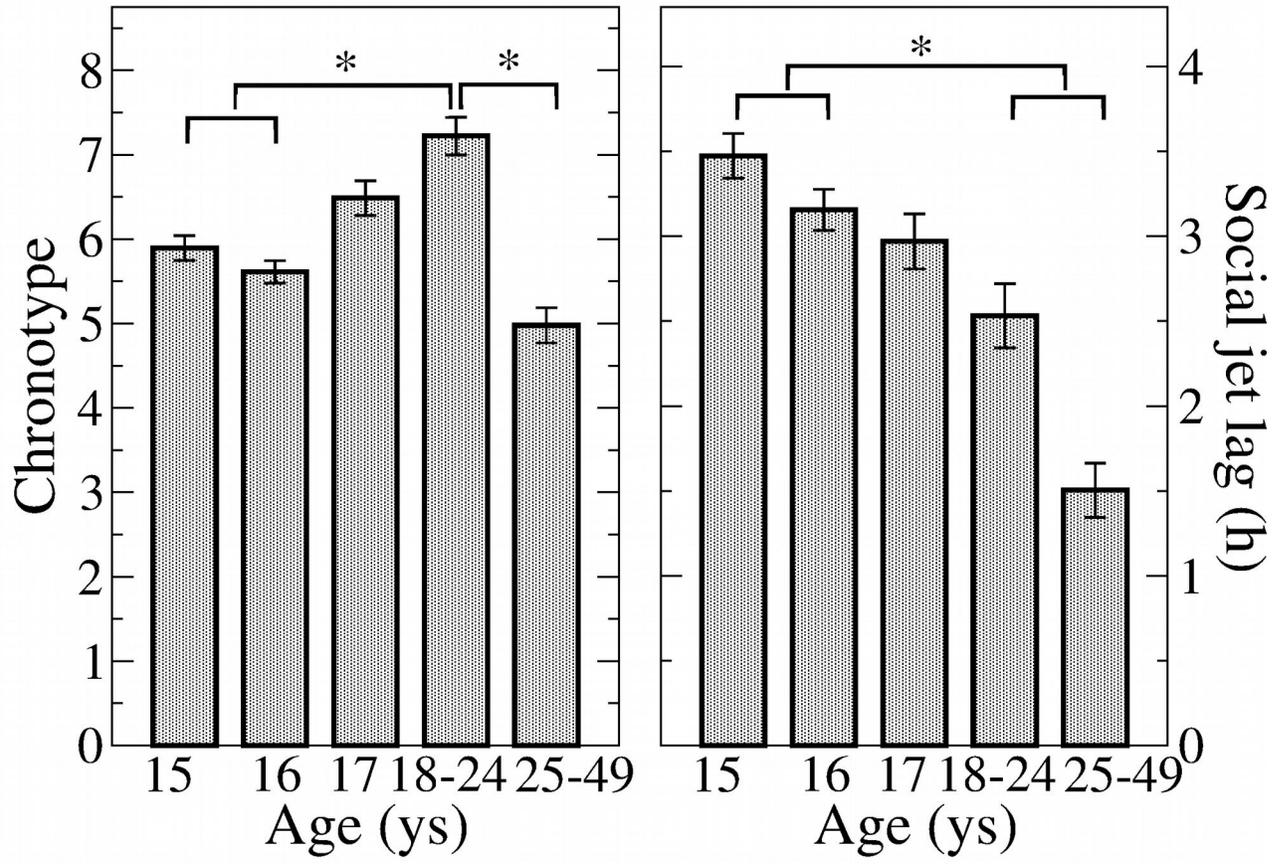

**Figure 9.**

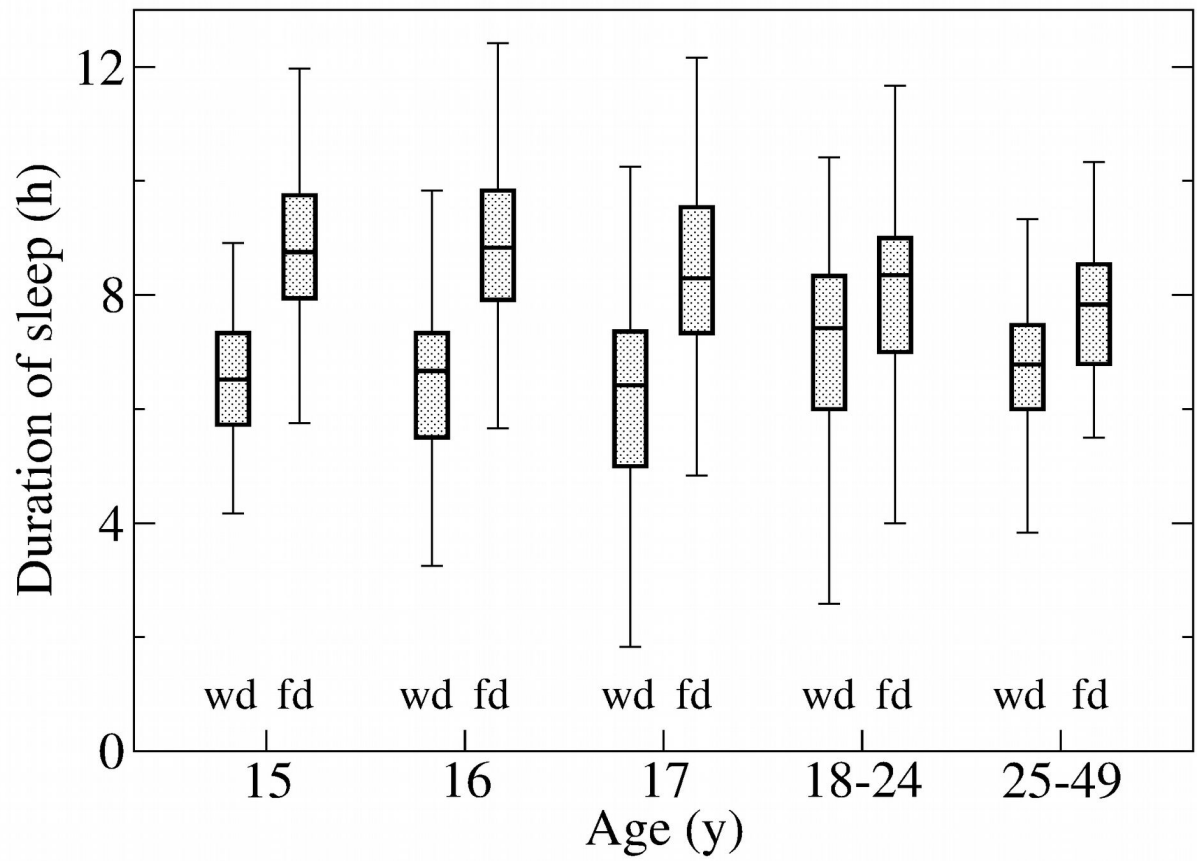

**Figure 10.**

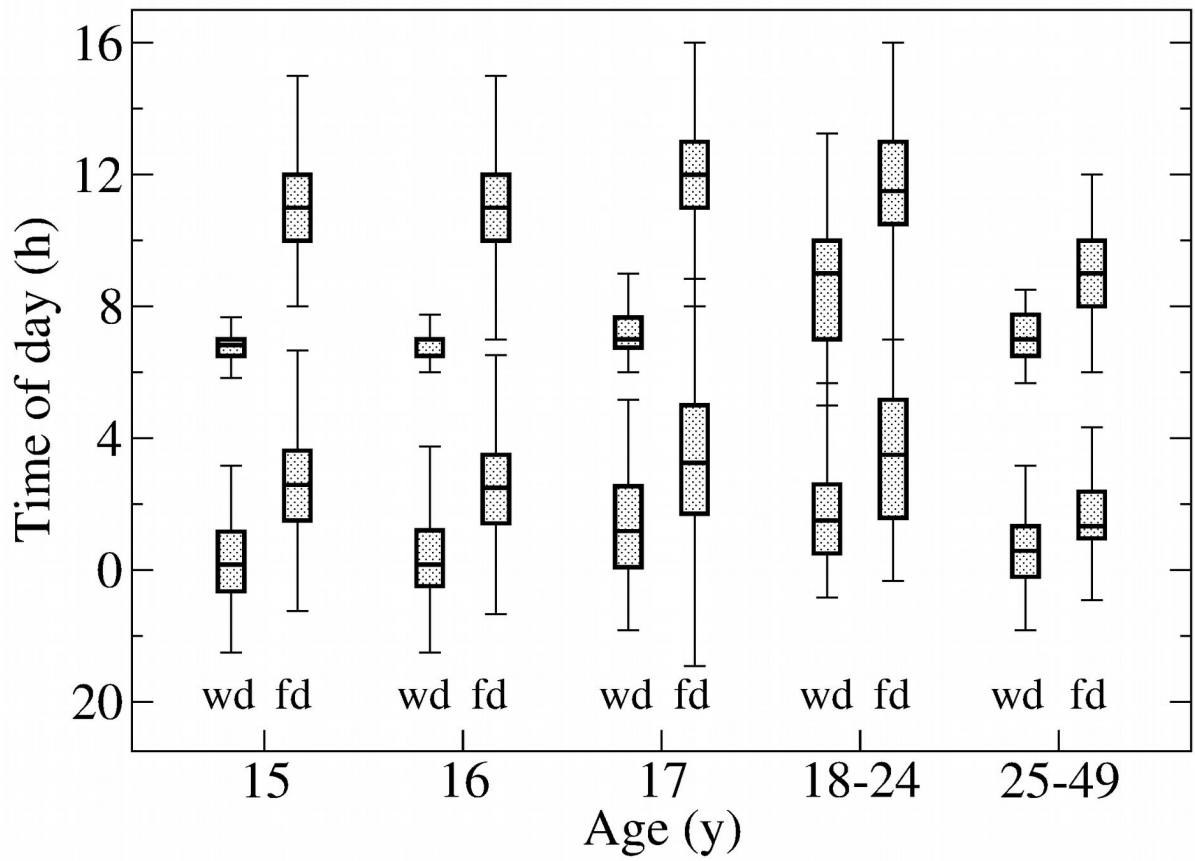